\documentclass{article}
\usepackage{graphicx}
\usepackage[final, nonatbib]{neurips_2023_ml4ps}
\usepackage[numbers,sort&compress]{natbib} 

\usepackage{multirow}
\usepackage{placeins}
\usepackage{siunitx}
\usepackage{tikz}
\usepackage{subcaption}
\usepackage{hyperref}
\usepackage{algorithm}
\usepackage{algpseudocode}
\usepackage{amsmath}
\usepackage{booktabs}

\title{Active learning meets fractal decision boundaries: a cautionary tale from the Sitnikov three-body problem}
\author{\small{Nicolas Payot}\\
\small{Département de Physique, Université de Montréal}\\
\small{Mila - Quebec Artificial Intelligence Institute}\\
\small{Ciela - Montreal Institute for Astrophysical Data Analysis and Machine Learning}\\
\vspace{-0.3in}
\AND
\small{Mario Pasquato}\\
\small{Département de Physique, Université de Montréal}\\ \small{Mila - Quebec Artificial Intelligence Institute}\\
\small{Ciela - Montreal Institute for Astrophysical Data Analysis and Machine Learning}\\
\small{Dipartimento di Fisica e Astronomia, Università di Padova}\\
\small{Vicolo dell'Osservatorio 5, I-35122, Padova, Italy}\\
\vspace{-0.3in}
\And
\small{Alessandro Alberto Trani}\\
\small{Niels Bohr Institute, Copenhagen, Denmark}\\ 
\small{Research Center for the Early Universe, School of Science, The University of Tokyo, Tokyo, Japan}\\ 
\small{Okinawa Institute of Science and Technology, Okinawa, Japan}\\
\vspace{-0.3in}
\And
\small{Yashar Hezaveh}\\
\small{Département de Physique, Université de Montréal}\\ \small{Mila - Quebec Artificial Intelligence Institute}\\
\small{Ciela - Montreal Institute for Astrophysical Data Analysis and Machine Learning}\\
\vspace{-0.3in}
\And
\small{Laurence Perreault-Levasseur}\\
\small{Département de Physique, Université de Montréal}\\ \small{Mila - Quebec Artificial Intelligence Institute}\\
\small{Ciela - Montreal Institute for Astrophysical Data Analysis and Machine Learning}\\
\vspace{-0.3in}
}

\begin{document}

\maketitle

\begin{abstract}
Chaotic systems such as the gravitational N-body problem are ubiquitous in astronomy. Machine learning (ML) is increasingly deployed to predict the evolution of such systems, e.g. with the goal of speeding up simulations. Strategies such as active Learning (AL) are a natural choice to optimize ML training. Here we showcase an AL failure when predicting the stability of the Sitnikov three-body problem, the simplest case of N-body problem displaying chaotic behavior. We link this failure to the fractal nature of our classification problem's decision boundary. This is a potential pitfall in optimizing large sets of N-body simulations via AL in the context of star cluster physics, galactic dynamics, or cosmology.
\end{abstract}

\section{Introduction}
The gravitational N-body problem lies at the core of computational astronomy. Since the work of Laplace and Lagrange in the 18th century, the stability of the Solar system has been a prominent issue, spurring research to this day. More recently, the gravitational N-body problem is being solved numerically to model systems ranging from icy fragments within the rings of Saturn \citep[][]{2001Icar..154..296D} up to the cosmological scale \citep[][]{2015MNRAS.454...83W}. Astronomers increasingly apply ML to predict properties of gravitational N-body and related systems, including their dynamical evolution in the chaotic regime \citep[][]{2020MNRAS.498.2957C, 2020MNRAS.494.2465B, 2021arXiv211115631C, 2022arXiv220202306L, 2022AcAau.193..710Y, 2022JPhCS2243a2010I, 2022AdSpR..69.2865Y, 2022NatSR..12.1890C, 2022MNRAS.511.2218L, 2022ApJ...938...18L, 2023MNRAS.524.1374L}. 

Predicting the evolution of a chaotic system presents challenges due to sensitive dependence on initial conditions \citep[see e.g.][]{2020JCoPh.41809629H}, to the point that chaotic benchmarks were proposed for evaluating data-driven forecasting models \citep[][]{2021arXiv211005266G}. Determining the limits of applicability of ML to these systems is thus crucial for astronomy. Here we focus on a narrower question: are techniques such a AL always helpful in training ML models on the gravitational N-body problem in the chaotic regime?

In AL, a model queries the data it deems most informative from an unlabeled pool, selectively requesting labeling with the goal of using fewer labeled samples. This is beneficial when labeled data is costly, e.g. when labelling requires running expensive simulations. AL is becoming increasingly more popular in astronomy, with an early focus on observational surveys \citep[][]{2022A&A...663A..13L}.
We thus set out to test AL in the simplest N-body setting, a restricted version of the three-body problem known a the Sitnikov problem.
This is arguably the simplest case of N-body problem capable of displaying chaotic behavior. The space of its initial conditions is two-dimensional, allowing easy visualization on the plane, and the boundary between stable and unstable systems is indeed fractal.

We train different deep learning models with the goal of predicting the stability of Sitnikov's problem from its initial conditions, with and without AL. We find that our model's performance is not improved, all else equal, by AL. We attribute this to the fact that the decision boundary of our problem is fractal, which results in our AL strategy wasting queries to probe it, resulting in a suboptimal sampling of feature space.

\section{The Sitnikov problem} 

The Sitnikov three-body problem is a special case of the three-body problem wherein two masses orbit around their common barycenter while a third mass undergoes oscillations along the z axis. For eccentricities  of the orbit of the first and the second body greater than $0$, it exhibits chaotic behaviour, while the case with $e=0$ is integrable \citep{macmillan_integrable_1911, sitnikov_existence_1961, hagel_high_2005}.

We studied the $e=0.5$ case and we selected $m_1$ and $m_2$ such that $G(m_1 + m_2) = 1$ and $m_1$ = $m_2$. The motion of the third body in the Sitnikov problem is then governed by the following differential equations:
\begin{align}
    \frac{dz}{dt} &= v_z \\ 
    \frac{dv_z}{dt} &= - \frac{G(m_1+m_2)z}{({r(t)^2 + z^2})^{3/2}}
\end{align}
Here, $z$ and $v_z$ are the position and velocity of the third body on the z-axis, while $m_i$ is the mass of the $i^{th}$ body. $r(t)$ represents the distance between either the first or the second body and their center of mass. The oscillations are independent of the mass of the oscillating body, which is considered massless \citep{hagel_high_2005}.

This problem involves only two parameters: the initial position of the oscillating body, denoted as $z_0$, and its initial velocity along the z-axis, denoted as $v_{z_0}$. Changing the initial position of the two orbiting masses is equivalent to altering these two parameters. Consequently, all our simulations commence with the two massive bodies positioned at their respective apocenters.

We implemented our own Runge-Kutta-Felhberg solver of order 7(8) (RK7(8)) \footnote{\href{https://github.com/rouzib/SitnikovSolver}{https://github.com/rouzib/SitnikovSolver}} to minimize the numerical error of our simulations \citep{fehlberg_classical_1968}. We employed a variable time step to ensure that the error remained below $\mathcal{O}(10^{-12})$ as it was demonstrated to preserve the analytical solution \citep{dvorak_celestial_2013}.

A grid of values was computed with $500$ values of $v_{z_0}$ for each of the $200$ $z_0$ samples, resulting in a total of $100\,000$ simulations:
\begin{align*}
    z_0 &\in \left[-1, 1\right] \\
    v_{z_0} &\in \left[0, 3.5\right]
\end{align*}

The simulations were conducted for $100$ orbits of the massive bodies after which they were considered stable. If the oscillations of the massless particle were found to be unstable based on straightforward criteria, the simulation was halted, and the initial conditions were classified as unstable. We define instability as an orbit in which the massless body does not pass through the center of mass again. The criteria used to ascertain instability are as follows:

Essentially, if the body is significantly distant from the center of mass and its acceleration is insufficient to complete another oscillation, it is classified as unstable. A similar criterion applies in cases where $z < -10$. These values were selected empirically and validated on a smaller subset of data to confirm their accuracy.

\section{AL strategy}

We began by implementing the classification uncertainty as the AL strategy function. Since the classifier predicts $1$ if the simulation is stable or $0$ if the simulation breaks up before the end of the simulated time, the maximum uncertainty for classification is $0.5$. The classification uncertainty is defined as follows:
\begin{equation}
    U(x) = 2 \big|0.5 - P(\hat{x}|x)\big|
\end{equation}
where $x$ represents the simulation under study, and $\hat{x}$ is the classifier's prediction \citep{Settles2009ActiveLL}. This formulation assigns an uncertainty value close to 0 for the most uncertain simulations and 1 for the most certain ones. While this strategy is straightforward and relatively computationally efficient - as all samples need to be passed through the network - it solely considers the current training model. Consequently, this method does not take into account the intrinsic data characteristics and largely overlooks the data space. Unfortunately, this approach of uncertainty sampling can potentially lead to overfitting when utilizing simulation data, which was observed in this work \citep{yang_multi-class_2015}.

Subsequently, another strategy was implemented: the ranked batch-mode sampling, where space exploration is prioritized, and uncertainty sampling is applied subsequently \citep{cardoso_ranked_2017}. The scoring function is defined as:
\begin{equation}
    \text{score}(x) = \alpha(1-S(x, Y)) + (1 - \alpha)U(x)
\end{equation}
\begin{equation}
    \alpha = \max\left(\frac{N_X}{N_Y + N_X}c, 1\right)
\end{equation}
Here, $Y$ represents the pool of labeled data or the labeled dataset, $X$ is the set of unlabeled data, and $N_i$ denotes the number of samples in a dataset. The parameter $\alpha$ acts as a weighted control that decides whether the sampling process emphasizes space exploration or relies on uncertainty sampling. Initially, $\alpha$ is set to a higher value, but as more data is labeled using the AL strategy, the unlabeled dataset decreases in size, causing $\alpha$ to decrease. When $\alpha$ approaches $1$, the strategy prioritizes unexplored regions of the space for selecting new samples, while $\alpha$ close to $0$ signifies reliance on model uncertainty. The scalar parameter $c$ is accessible to the user and determines the rate at which the transition between the two strategies occurs, making it a new hyper-parameter. The similarity function $S(x,Y)$ measures the similarity between $x$ - the sample under study - and $y$ - the labeled dataset. The value returned by $S(x,Y)$ should fall between $0$ and $1$, with $1$ indicating that the sample $x$ lies in a completely unexplored portion of the space, and $0$ denoting that it coincides with another sample. For instance, cosine similarity can be employed as a suitable choice \citep{cardoso_ranked_2017}. The sample with the minimum score is then selected for labeling, added to the labeled dataset, and removed from the unlabeled dataset. This process is repeated for the required number of queries.

\section{Results}
\vspace{-0.08in}
\begin{figure}[ht]
    \centering
    \includegraphics[width=\textwidth]{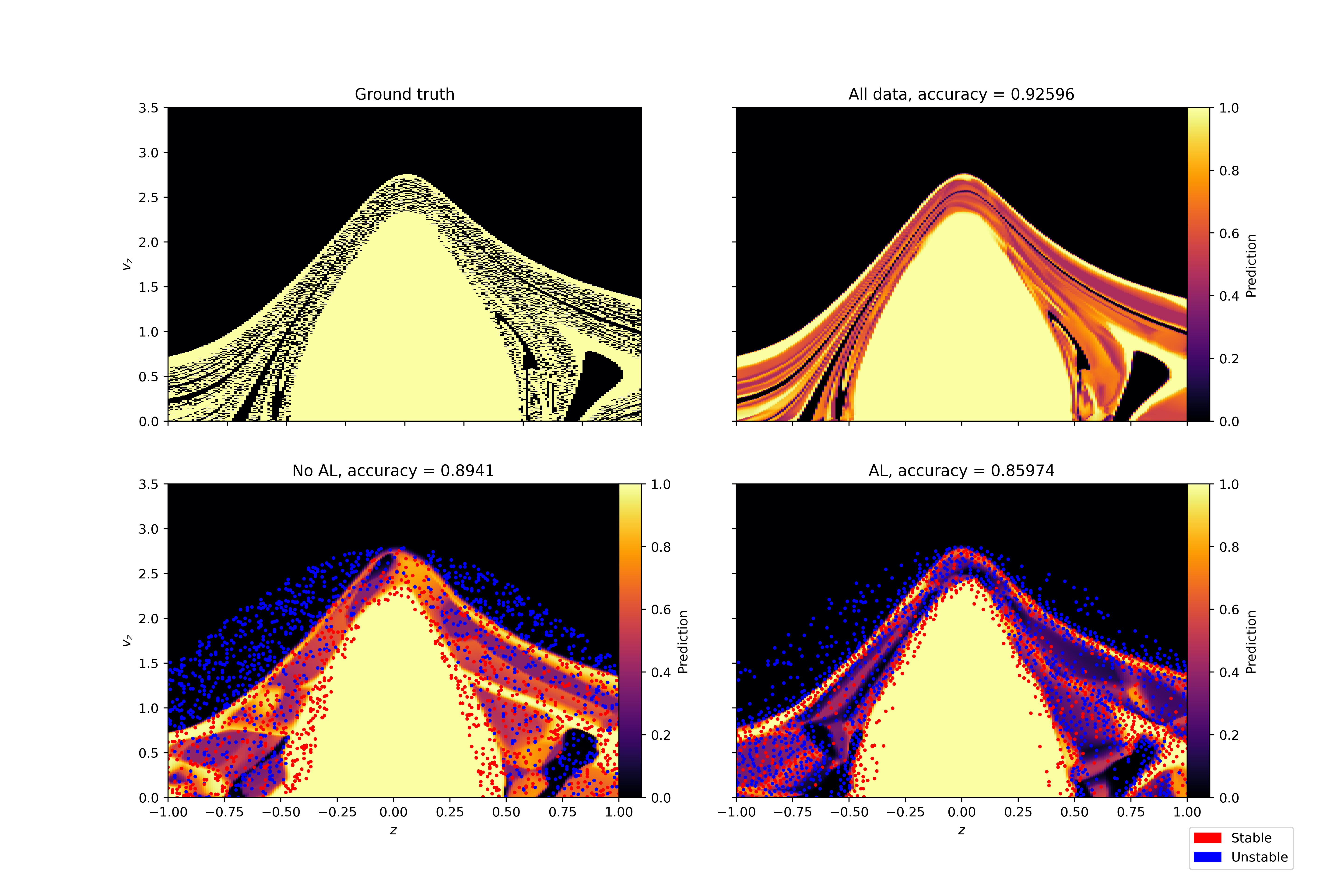}
    \caption{Top row: Ground truth and a model trained on all the data, achieving an accuracy of $92.596\%$. The ground truth is limited by the number of simulations performed. Bottom row: A model trained without AL, which achieved an accuracy of $89.410\%$, and a model trained with AL, with an accuracy of $85.974\%$. The blue and red points respectively represent unstable and stable simulations used during training.}
    \label{fig:decision}
\end{figure}

The neural network employed is a straightforward fully-connected network comprising seven layers. We conducted tests with this model on all our simulations to confirm its capability to effectively learn the spatial information [see figure \ref{fig:decision}]. The neural network generated a single output value: $1$ if it predicted that the initial conditions would lead to a stable system or $0$ if not. Since most of the data could be easily classified as stable or unstable, the dataset was pruned to eliminate these trivial cases. Out of the original $100\,000$ simulations, only $46\,502$ fell within the selected criteria. For AL, $400\,000$ potential initial conditions were generated from a uniform distribution and selected using the same criteria as for the dataset. The remaining $278\,980$ initial conditions were designated for AL sampling. When AL was not employed, the same number of simulations as would be seen in total by the AL strategy were selected from the dataset.

We utilized a learning rate of $0.001$ and set $c$ to $0.5$, to facilitate exploration of the parameter space, with a primary focus on modeling uncertainty. We examined a range of values for $c$ from $0$ to $1$, determining that $c=0.5$ yielded the most favorable results. All models underwent a training process that spanned $400$ epochs, utilizing a NVIDIA A100. In the context of our AL approach, the neural network began with an initial dataset of $500$ simulations. As the training progressed, we incrementally added $10$ additional simulations from the initial condition dataset during each epoch, specifically from epochs $50$ to $200$, accumulating a total of $1500$ new simulations.

\begin{figure}[ht]
    \centering
    \includegraphics[width=0.8\textwidth]{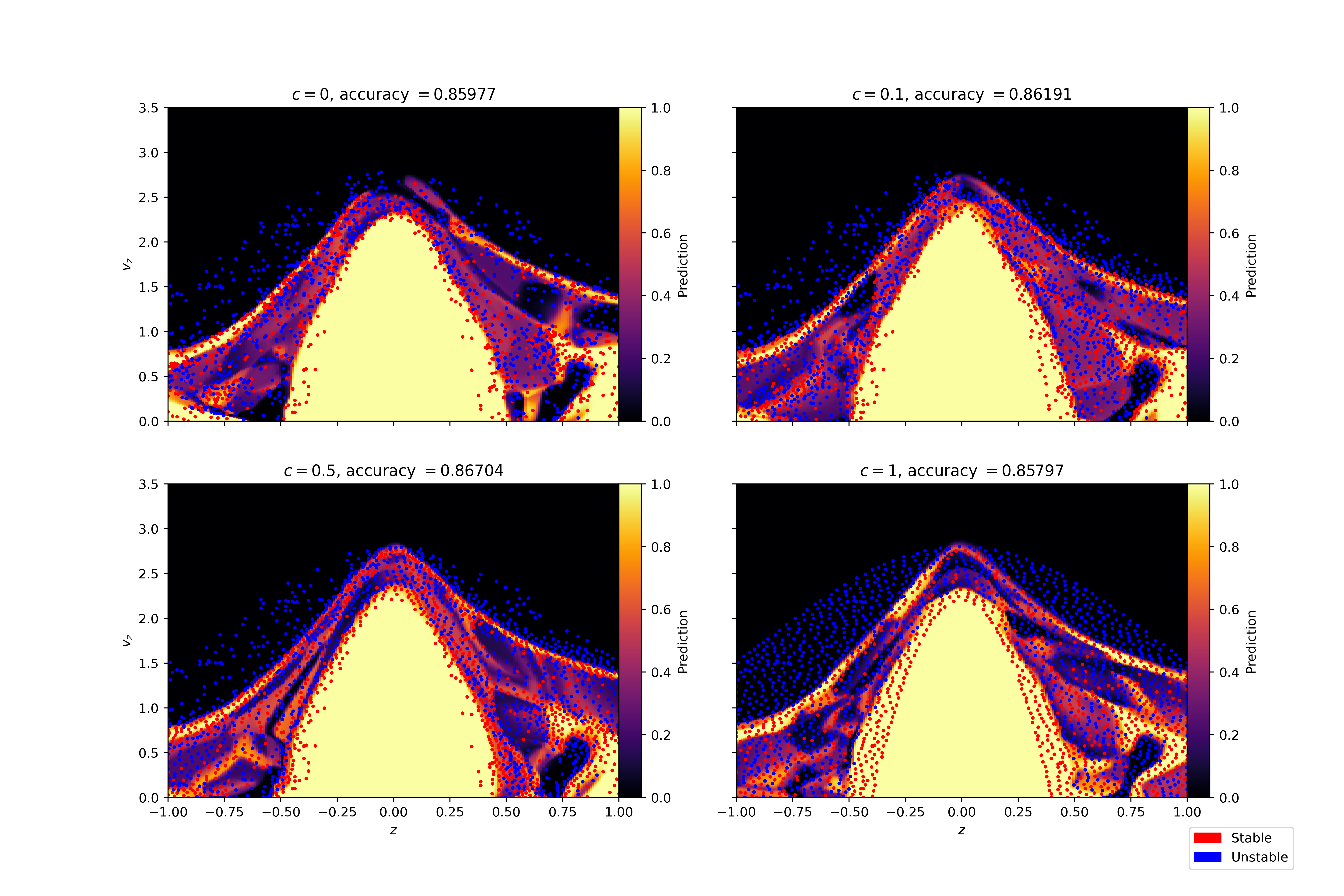}
    \caption{Models trained using the same parameters as previously stated, but with different values for $c$ ranging from $0$ to $1$. The model achieving the best accuracy is the one trained with $c=0.5$. The blue and red points respectively represent unstable and stable simulations used during training.}
    \label{fig:c}
\end{figure}

\begin{figure}[ht]
    \centering
    \includegraphics[width=0.5\textwidth]{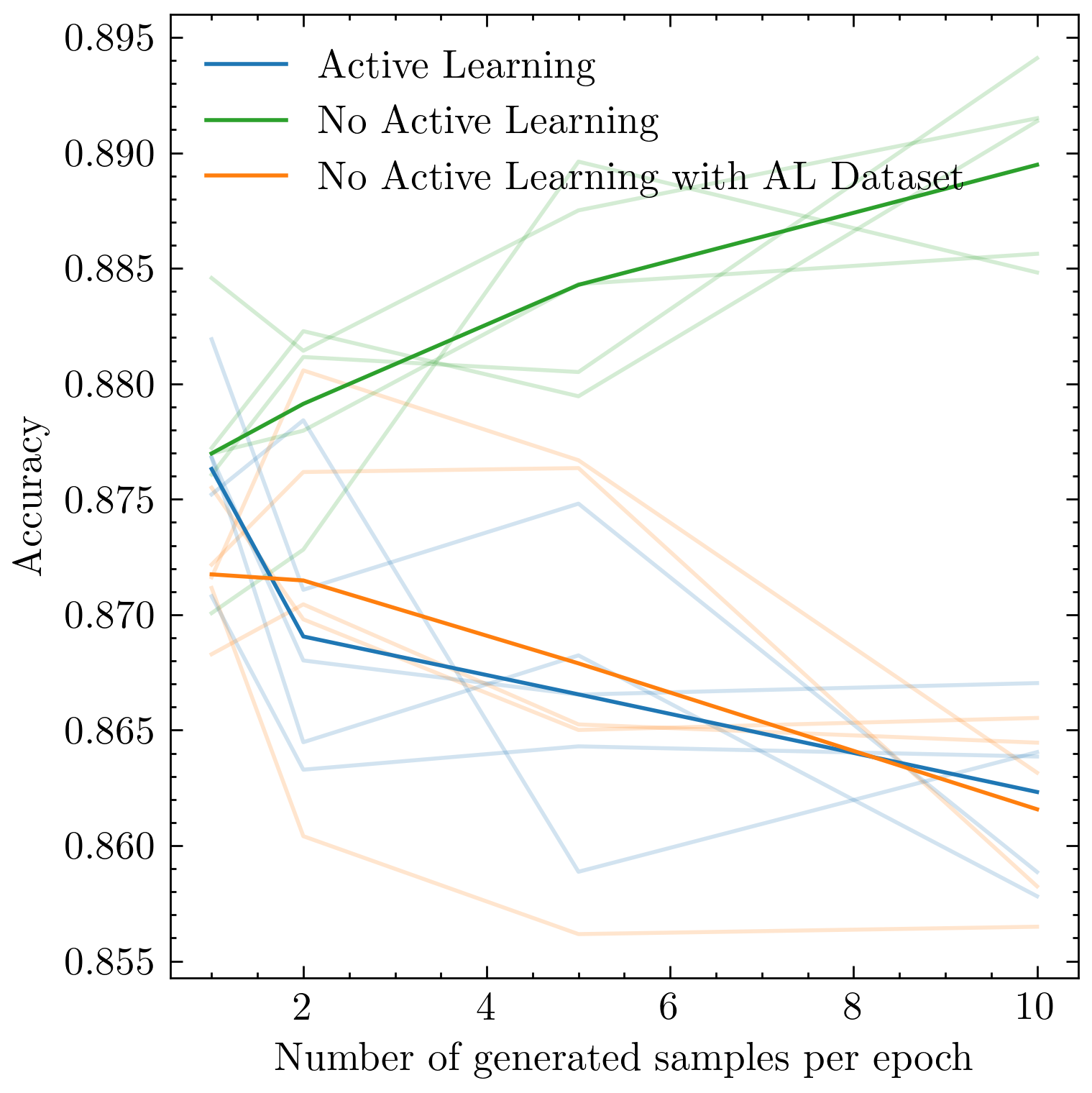}
    \caption{Comparison of model accuracies trained with AL and without AL, against the number of generated samples per epoch. The 'No Active Learning with AL Dataset' curve represents a model trained on a dataset selected through AL but without the AL strategy during training. The accuracy values represent averages over multiple training runs with different random seeds to ensure reliability of the performance measure.}
    \label{fig:accuracy}
\end{figure}

Incorporating AL significantly reduces the number of simulations sampled in the trivial cases, as expected. It places a focus on boundary cases where predicting the stability of a simulation is more challenging. However, a drawback arises due to the fractal-shaped boundaries in the Sitnikov problem, as documented in prior research \citep{kovacs_structure_2007, dvorak_numerical_1993}. The model never truly converges to a satisfactory result because in the chaotic regime arbitrarily small perturbations in initial conditions can lead to different outcomes.

Increasing $c$ pushes the model to explore the space more extensively. However, this exploration eventually essentially resembles random sampling. As the model begins to prioritize uncertainty sampling, it zooms in on these intricate and ill-defined boundaries. This excessive focus on narrow boundary regions results in the neglect of other parts of the space, leading to poor generalization. This behavior is depicted in Figure~\ref{fig:c}, which illustrates the impact of different values of $c$ on the model's ability to generalize and its performance.

We also investigated the model's performance when trained on the complete dataset gathered via the AL process, as opposed to employing random sampling. This approach was intended to ascertain if the simulations selected through AL were inherently the source of the poor performance. The outcomes revealed that the model trained on the AL-derived dataset demonstrated performance closely aligned with that of the model trained using AL itself [see figure \ref{fig:accuracy}].

\section{Conclusions}
We used the Sitnikov restricted three-body problem to compare the performance achieved by a neural network trained to predict stability using an AL approach versus random sampling. We found that AL does not result in improved performance. We attribute this failure to the fractal nature of the decision boundary for our problem. Real life applications of AL to optimize sets of gravitational N-body simulations are likely to face the same issue. Our results are therefore valuable as a cautionary tale for astronomers planning such an undertaking.

\newpage
\begin{ack}
M. P. acknowledges financial support from the European Union’s Horizon 2020 research and innovation program under the Marie Sk\l{}odowska-Curie grant agreement No. 896248. A.A.T. acknowledges support from the European Union’s Horizon 2020 and Horizon Europe research and innovation programs under the Marie Sk\l{}odowska-Curie grant agreements no. 847523 and 101103134.

This work is supported by the Simons Collaboration on “Learning the Universe". The Flatiron Institute is supported by the Simons Foundation. The work is in part supported by computational resources provided by Calcul Quebec and the Digital Research Alliance of Canada. Y.H. and L.P. acknowledge support from the Canada Research Chairs Program, the National Sciences and Engineering Council of Canada through grants RGPIN-2020- 05073 and 05102, and the Fonds de recherche du Québec through grants 2022-NC-301305 and 300397. P.L acknowledges support from the Simons Foundation.

\end{ack}

\bibliographystyle{unsrt}
\bibliography{refs}

\begin{thebibliography}{10}

\bibitem{2001Icar..154..296D}
Hiroshi {Daisaka}, Hidekazu {Tanaka}, and Shigeru {Ida}.
\newblock {Viscosity in a Dense Planetary Ring with Self-Gravitating Particles}.
\newblock {\em Icarus}, 154(2):296--312, December 2001.

\bibitem{2015MNRAS.454...83W}
Liang {Wang}, Aaron~A. {Dutton}, Gregory~S. {Stinson}, Andrea~V. {Macci{\`o}}, Camilla {Penzo}, Xi~{Kang}, Ben~W. {Keller}, and James {Wadsley}.
\newblock {NIHAO project - I. Reproducing the inefficiency of galaxy formation across cosmic time with a large sample of cosmological hydrodynamical simulations}.
\newblock {\em Monthly Notices of The Royal Astronomical Society}, 454(1):83--94, November 2015.

\bibitem{2020MNRAS.498.2957C}
T.~A.~F. {Comerford} and R.~G. {Izzard}.
\newblock {Estimating the outcomes of common envelope evolution in triple stellar systems}.
\newblock {\em Monthly Notices of The Royal Astronomical Society}, 498(2):2957--2967, October 2020.

\bibitem{2020MNRAS.494.2465B}
Philip~G. {Breen}, Christopher~N. {Foley}, Tjarda {Boekholt}, and Simon {Portegies Zwart}.
\newblock {Newton versus the machine: solving the chaotic three-body problem using deep neural networks}.
\newblock {\em Monthly Notices of The Royal Astronomical Society}, 494(2):2465--2470, May 2020.

\bibitem{2021arXiv211115631C}
Maxwell~X. {Cai}, Simon {Portegies Zwart}, and Damian {Podareanu}.
\newblock {Neural Symplectic Integrator with Hamiltonian Inductive Bias for the Gravitational $N$-body Problem}.
\newblock {\em arXiv e-prints}, page arXiv:2111.15631, November 2021.

\bibitem{2022arXiv220202306L}
Pablo {Lemos}, Niall {Jeffrey}, Miles {Cranmer}, Shirley {Ho}, and Peter {Battaglia}.
\newblock {Rediscovering orbital mechanics with machine learning}.
\newblock {\em arXiv e-prints}, page arXiv:2202.02306, February 2022.

\bibitem{2022AcAau.193..710Y}
Hongwei {Yang}, Jiumei {Yan}, and Shuang {Li}.
\newblock {Fast computation of the Jovian-moon three-body flyby map based on artificial neural networks}.
\newblock {\em Acta Astronautica}, 193:710--720, April 2022.

\bibitem{2022JPhCS2243a2010I}
M.~I. {Ikhsan} and M.~I. {Arifyanto}.
\newblock {Exploring multi-planet system wasp-148 using n-body simulation and deep learning}.
\newblock In {\em Journal of Physics Conference Series}, volume 2243 of {\em Journal of Physics Conference Series}, page 012010, June 2022.

\bibitem{2022AdSpR..69.2865Y}
Jiumei {Yan}, Hongwei {Yang}, and Shuang {Li}.
\newblock {ANN-based method for fast optimization of Jovian-moon gravity-assisted trajectories in CR3BP}.
\newblock {\em Advances in Space Research}, 69(7):2865--2882, April 2022.

\bibitem{2022NatSR..12.1890C}
Alessandra {Celletti}, Catalin {Gales}, Victor {Rodriguez-Fernandez}, and Massimiliano {Vasile}.
\newblock {Classification of regular and chaotic motions in Hamiltonian systems with deep learning}.
\newblock {\em Scientific Reports}, 12:1890, January 2022.

\bibitem{2022MNRAS.511.2218L}
Xin {Li}, Jian {Li}, Zhihong~Jeff {Xia}, and Nikolaos {Georgakarakos}.
\newblock {Machine-learning prediction for mean motion resonance behaviour - The planar case}.
\newblock {\em Monthly Notices of The Royal Astronomical Society}, 511(2):2218--2228, April 2022.

\bibitem{2022ApJ...938...18L}
Florian {Lalande} and Alessandro~Alberto {Trani}.
\newblock {Predicting the Stability of Hierarchical Triple Systems with Convolutional Neural Networks}.
\newblock {\em Astrophysical Journal}, 938(1):18, October 2022.

\bibitem{2023MNRAS.524.1374L}
Xin {Li}, Jian {Li}, Zhihong~Jeff {Xia}, and Nikolaos {Georgakarakos}.
\newblock {Large-step neural network for learning the symplectic evolution from partitioned data}.
\newblock {\em Monthly Notices of The Royal Astronomical Society}, 524(1):1374--1385, September 2023.

\bibitem{2020JCoPh.41809629H}
Tianli {Hu} and Shijun {Liao}.
\newblock {On the risks of using double precision in numerical simulations of spatio-temporal chaos}.
\newblock {\em Journal of Computational Physics}, 418:109629, October 2020.

\bibitem{2021arXiv211005266G}
William {Gilpin}.
\newblock {Chaos as an interpretable benchmark for forecasting and data-driven modelling}.
\newblock {\em arXiv e-prints}, page arXiv:2110.05266, October 2021.

\bibitem{2022A&A...663A..13L}
M.~{Leoni}, E.~E.~O. {Ishida}, J.~{Peloton}, and A.~{M{\"o}ller}.
\newblock {Fink: Early supernovae Ia classification using active learning}.
\newblock {\em Astronomy and Astrophysics}, 663:A13, July 2022.

\bibitem{macmillan_integrable_1911}
W.~D. MacMillan.
\newblock An integrable case in the restricted problem of three bodies.
\newblock {\em The Astronomical Journal}, 27:11, May 1911.

\bibitem{sitnikov_existence_1961}
K.~Sitnikov.
\newblock The {Existence} of {Oscillatory} {Motions} in the {Three}-{Body} {Problem}.
\newblock {\em Soviet Physics Doklady}, 5:647, January 1961.

\bibitem{hagel_high_2005}
Johannes Hagel and Christoph Lhotka.
\newblock A {High} {Order} {Perturbation} {Analysis} of the {Sitnikov} {Problem}.
\newblock {\em Celestial Mech Dyn Astr}, 93(1-4):201--228, September 2005.

\bibitem{fehlberg_classical_1968}
Erwin Fehlberg.
\newblock {\em Classical {Fifth}-, {Sixth}-, {Seventh}-, and {Eighth}-order {Runge}-{Kutta} {Formulas} with {Stepsize} {Control}}.
\newblock National Aeronautics and Space Administration, 1968.

\bibitem{dvorak_celestial_2013}
Rudolf Dvorak and Christoph Lhotka.
\newblock {\em Celestial {Dynamics}: {Chaoticity} and {Dynamics} of {Celestial} {Systems}}.
\newblock Wiley, 1 edition, April 2013.

\bibitem{Settles2009ActiveLL}
Burr Settles.
\newblock Active learning literature survey.
\newblock In {\em Active Learning Literature Survey}, 2009.

\bibitem{yang_multi-class_2015}
Yi~Yang, Zhigang Ma, Feiping Nie, Xiaojun Chang, and Alexander~G. Hauptmann.
\newblock Multi-{Class} {Active} {Learning} by {Uncertainty} {Sampling} with {Diversity} {Maximization}.
\newblock {\em Int J Comput Vis}, 113(2):113--127, June 2015.

\bibitem{cardoso_ranked_2017}
Thiago~N.C. Cardoso, Rodrigo~M. Silva, Sérgio Canuto, Mirella~M. Moro, and Marcos~A. Gonçalves.
\newblock Ranked batch-mode active learning.
\newblock {\em Information Sciences}, 379:313--337, February 2017.

\bibitem{kovacs_structure_2007}
T.~Kovács and B.~Érdi.
\newblock The structure of the extended phase space of the {Sitnikov} problem.
\newblock {\em Astron. Nachr.}, 328(8):801--804, October 2007.

\bibitem{dvorak_numerical_1993}
R.~Dvorak.
\newblock Numerical {Results} to the {Sitnikov}-{Problem}.
\newblock {\em Celestial Mechanics and Dynamical Astronomy}, 56:71--80, May 1993.
\newblock ADS Bibcode: 1993CeMDA..56...71D.

\bibitem{trani_keplerian_2019}
Alessandro~A. Trani, Mario Spera, Nathan W.~C. Leigh, and Michiko~S. Fujii.
\newblock The keplerian three-body encounter. {II}. comparisons with isolated encounters and impact on gravitational wave merger timescales.
\newblock {\em {ApJ}}, 885(2):135, 2019.

\bibitem{trani_keplerian_2019-1}
Alessandro~A. Trani, Michiko~S. Fujii, and Mario Spera.
\newblock The keplerian three-body encounter. i. insights on the origin of the s-stars and the g-objects in the galactic center.
\newblock {\em {ApJ}}, 875(1):42, 2019.

\bibitem{lalande_predicting_2022}
Florian Lalande and Alessandro~Alberto Trani.
\newblock Predicting the stability of hierarchical triple systems with convolutional neural networks.
\newblock {\em {ApJ}}, 938(1):18, 2022.

\end{thebibliography}

\newpage
\section{Appendix}
\FloatBarrier
\begin{algorithm}
\caption{Score pseudo code}\label{alg:scoreFast}
\begin{algorithmic}
\Function{score}{$x$, $Y$, $c$, $u$} \Comment{$x$ is the total data set, $Y$ is the labeled or training data set, $u$ is the uncertainty}
    \State $\alpha \gets \frac{len(x)}{len(x) + len(Y)} * c$
    \State $term_1 \gets \alpha * (1 - S(x, Y))$
    \State $term_2 \gets (1-\alpha) * u$
    \State \Return $term_1 + term_2$
\EndFunction
\end{algorithmic}
\end{algorithm}

\begin{algorithm}
\caption{Uncertainty pseudo code}\label{alg:u}
\begin{algorithmic}
\Function{u}{$x$} \Comment{See Algorithm \ref{alg:queryFast} for usage}
   \State subdiv $\gets 64$ \Comment{Same as batch size}
   \State res $\gets$ [] \Comment{Will contain the uncertainties}
   \For{$i \text{ from } 0 \text{ to } \left(len(x) //\texttt{subdiv}+1\right)$}
        \State end $\gets \text{min}\left(len(x), (i+1)*\text{subdiv}\right)$
        \State predictions $\gets \text{model}\left(x[i*\text{subdiv}:\text{end}]\right)$
        \State predictions $\gets 2* \text{abs}\left(0.5 - \text{predictions}\right)$
        \State res $\gets$ Concatenate $\left(\text{res}, \text{ predictions}\right)$
    \EndFor
     \State \Return res
\EndFunction
\end{algorithmic}
\end{algorithm}

\begin{algorithm}
\caption{AL query pseudo code}\label{alg:queryFast}
\begin{algorithmic}
\Function{query}{$n$, $x$, $Y$, $c$} \Comment{$n$ is the amount of samples to query, $x$ is the total data set, $Y$ is the labeled or training data set}
    \State toLabel $\gets$ []
    \State $u \gets u(x)$ 
    \For{$i \text{ from } 0 \text{ to } n$}
    \State scores $\gets$ SCORE($x$, $Y$, $c$, $u$)
    \State idx $\gets$ argmin$($scores$)$
    \State toLabel $\gets$ Concatenate $\left(\text{toLabel}, \text{ x[idx]}\right)$
    \State Y $\gets$ Concatenate $\left(\text{Y}, \text{ x[idx]}\right)$
    \State x $\gets$ Concatenate $\left(\text{x[:idx]}, \text{ x[idx+1:]}\right)$
    \State $u$ $\gets$ Concatenate $\left(\text{$u$[:idx]}, \text{ $u$[idx+1:]}\right)$
    \EndFor
    \State \Return toLabel
\EndFunction
\end{algorithmic}
\end{algorithm}
\FloatBarrier

After \ref{alg:queryFast}, the new samples need to be labeled using an oracle. In our case, we run the simulations to see if they are stable or not.

\newpage

A neural network's capability to effectively classify three-body problems using time series data was demonstrated in \citep{2022ApJ...938...18L}. Building on this, we employed a simple neural network to the initial conditions of a generalized three-body simulation to predict stability, utilizing the same active learning scheme. Our investigation focused on a scenario where a star orbits a binary system. The simulations were carried out within the TSUNAMI framework \citep{trani_keplerian_2019, trani_keplerian_2019-1}, exploring a wide range of initial conditions given in Table~\ref{tab:SimParams}. For simplicity, all masses were assumed to be equal.

\begin{table}[!h]
    \centering
    \begin{tabular}{@{}cccl@{}}
    \toprule
    \multicolumn{1}{c}{\textbf{Category}} & \textbf{Parameter} & \textbf{Symbol} & \multicolumn{1}{c}{\textbf{Value}} \\
    \midrule
    \multirow{5}{*}{Binary} 
    & Semimajor axis & \(a_1\) & \(1\) \\ 
    & Eccentricity & \(e_1\) & \(U[0, 1]\) \\ 
    & Argument of periapsis & \(\omega_1\) & \(U[-\pi, \pi]\) \\ 
    & Longitude of the ascending node & \(\Omega_1\) & \(U[0, 2\pi]\) \\ 
    & Inclination & \(i_1\) & \(U[0, \pi]\) \\ 
    \midrule
    \multirow{3}{*}{Orbiting star} 
    & Semimajor axis & \(a_2\) & \(U[0, 3]\) \\ 
    & Eccentricity & \(e_2\) & \(U[0, 1]\) \\ 
    & Argument of periapsis & \(\omega_2\) & \(U[-\pi, \pi]\) \\
    \bottomrule
    \end{tabular}
    \caption{Orbital elements sampled for the generalized three-body simulations. $U[a, b]$ denotes a uniform distribution between the limits \(a\) and \(b\).}
    \label{tab:SimParams}
\end{table}

The distribution of the second semimajor axis, $a_2$, was chosen to balance the number of stable and unstable simulations. Orbital elements were sampled using the Latin hypercube sampling method, initially between $0$ and $1$, then scaled to the appropriate values. The inclination of the orbiting star and its longitude of the ascending node were set to $0$, as they merely result in a rotation of the plane in which the simulation is conducted, providing no additional information. The true anomaly of both the binary and the orbiting star was initialized at $0$.

Systems were simulated for a duration expressed in units of the binary's first complete orbit time divided by $2\pi$, denoted as $P_1$.
\begin{equation}
    P_1 = \sqrt{\frac{a_1^3}{G(m_1+m_2)}} = \sqrt{\frac{1}{2m_1G}}
\end{equation}
Simulations were retained only if the system remained intact for at least $500 P_1$. They were then continued until $10,000 P_1$ or a breakup occurred. Breakup was detected if the ratio of the two semimajor axes deviated from the initial conditions by more than $15\%$, a criterion also utilized in other classification tasks \citep{lalande_predicting_2022}. Orbital elements were recorded every $\pi P_1$ for the first $500 P_1$, provided no breakup occurred before this threshold. Simulations were deemed unstable if $e_1$ or $e_2$ fell outside the inclusive range of $0$ to $1$ exclusive.

\begin{figure}[ht]
    \centering
    \includegraphics[width=0.6\textwidth]{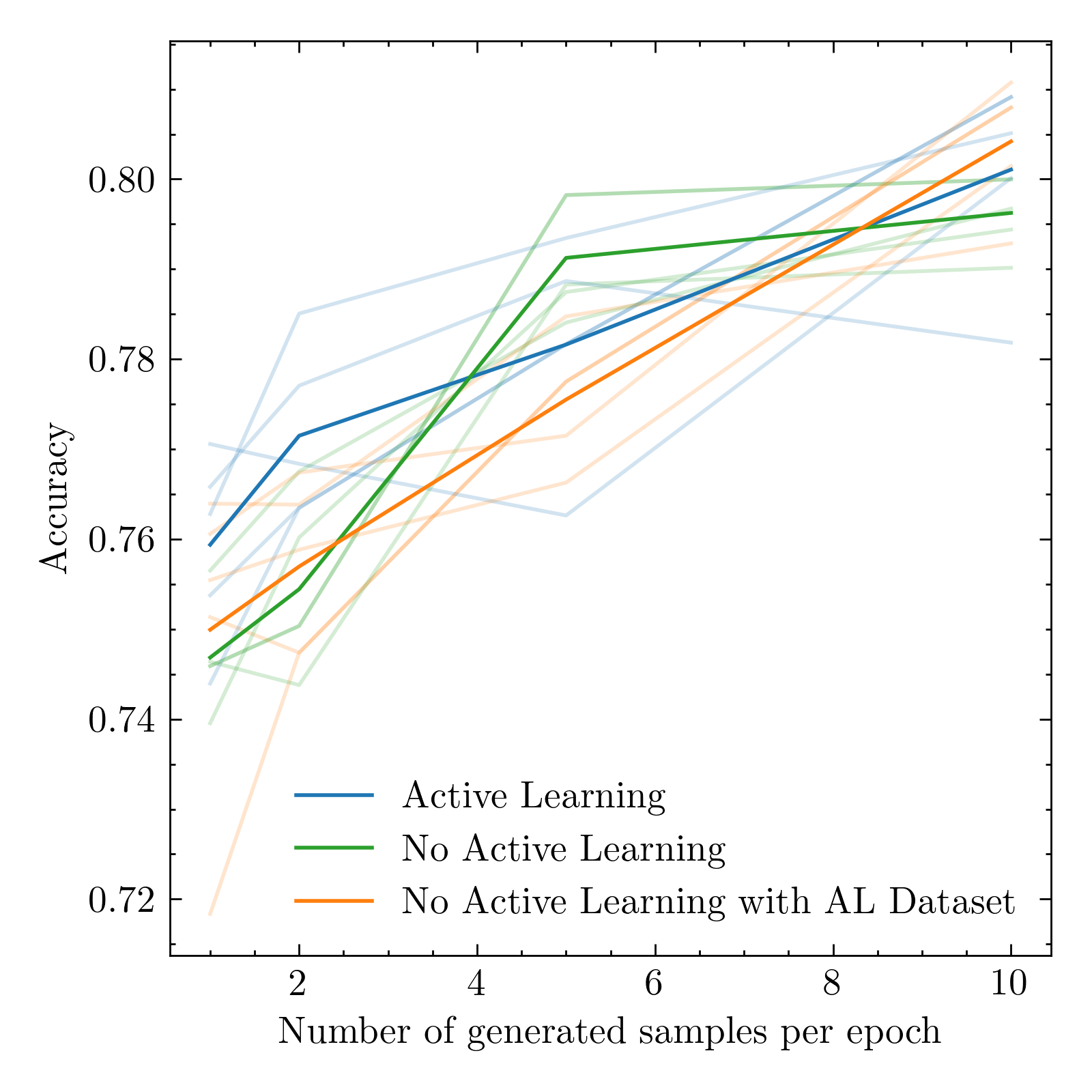}
    \caption{Comparison of model accuracies using Active Learning (AL) and without AL against the number of samples generated per epoch in the generalized three-body problem. The 'No Active Learning with AL Dataset' curve shows the accuracy of a model trained on an AL-selected dataset but without implementing AL during training. Accuracy values are averaged over multiple runs with varied random seeds to guarantee performance measure reliability.}
    \label{fig:accuracyGeneralized}
\end{figure}

A dataset comprising $48,262$ stable and $44,533$ unstable simulations was assembled to perform the active learning task. Unlike in the Sitnikov problem, applying the previously described AL scheme to this dataset did not exhibit the same drawbacks. We hypothesize that this difference arises from the smaller fractal and chaotic regions within the space of the generalized three-body problem. Consequently, the AL scheme is less likely to become ensnared in these complex areas. Additionally, the exploratory component of our AL strategy should aid in avoiding getting trapped in such smaller challenging regions, thereby not perturbing the overall effectiveness of the learning process.

\end{document}